# When is Good Good Enough? Context Factors for Good Remote Work of Agile Software Development Teams – The Otto Case

L. Rometsch[1], R. Wegner[1], F. Brusch[1], M. Neumann[1*], L. Linke[2]

[1] Hochschule Hannover, Faculty IV, Department of Business Informatics, Hannover, Germany
e-mail: {lisa.rometsch, richard.wegner, florian.brusch}@stud.hs-hannover.de,
michael.neumann@hs-hannover.de
[2]Otto GmbH & Co KG., E-Commerce Solutions & Technology, Hamburg, Germany
e-mail: lukas.linke@otto.de

**Abstract**—The Covid-19 pandemic led to several challenges in everybody working life. Many companies worldwide enabled comprehensive remote work settings for their employees. Agile Software Development Teams are affected by the switch to remote work as agile methods setting communication and collaboration in focus. The well-being and motivation of software engineers and developers, which impacting their performance, are influenced by specific context factors. This paper aims to analyze identify specific context factors for a good remote work setting. We designed a single case study at a German ecommerce company and conducted an experiment using a gamification approach including eight semi-structured interviews. Our results show, that the agile software development team members to their health. Furthermore, most the team members value the gamification approach to put more focus on physical activities and the health well-being. We discuss several practical implications and provide recommendations for other teams and companies.

## 1. INTRODUCTION

Agile methods are well-known approaches in the area of software development. The popularity of iterative, incremental, and lightweight approaches such as Scrum or XP has increased steadily over the past 20 years [1]. Agile methods are setting social aspects in focus [2]. Especially the communication and collaboration among the members of agile software development teams and further, the involvement and integration of stakeholders like customers or clients are of high importance for the success of agile methods [3].

The Covid 19 pandemic has led to various change processes worldwide over the past two years. Many countries and companies have defined and implemented extensive containment measures to reduce the spread of the virus. One of these measures is the switch to remote work. For instance, the German government has required companies to allow home-based work and let employees work from home when possible.

It is known that the quality of communication and collaboration can be negatively affected in distributed teams [4]. Agile software development teams have thus been affected by the shift to remote activity, as their daily collaboration has changed (e.g., [5], [6]). Various studies show effects on the social aspects of collaboration or even difficulties in onboarding and socializing new team members (e.g., [7-10]). Likewise, effects on specific agile practices that have a strong collaborative focus, such as pair programming, are described [11]. Various authors explain in studies that agile software development teams have reacted to the new circumstances and adapted their agile approach [7-12]. The effects on the performance of agile software development teams are described in several studies as both positive [7-10] and negative [6,13].

We know that well-being and motivation are highly important for the performance of software engineers and developers [14]. Good work for employees can be evaluated differently concerning the work situation, but depends on the design of certain context factors [15]. Thus, it is important to understand which context factors affect good work. One may assume that the work organization of remote and onsite work come with different context factors concerning the evaluation of good work.

This paper aims to analyze the context factors that enable good work in a remote work setting. In particular, we investigate which context factors influences the way of working to enable quality and a better performance of the agile software development team. We designed a single case study at Otto. Otto is a German trading and services company headquartered in Hamburg, which operates worldwide with around 52,000 employees in the business areas of e-commerce, retail, finance and logistics.

The two aims of our study are to define what constitutes good work and to identify the context factors for good work for a remote operating agile software development team at Otto. Thus, we defined the following research question:

**RQ:** *Which context factors are relevant for good work at Otto for a specific agile software development team working in a remote work setting and can be designed accordingly?*

The paper at hand is structured as follows: We provide the theoretical background including the fundamentals of good work and gamification in the second Section. In the third Section, we present our research design. We present the results of our study in Section 4. Before the paper closes with a conclusion and the limitations in Section 6, we discuss practical implications in Section 5.

## 2. THEORETICAL BACKGROUND

### 2.1. *Fundamentals of Good Work*

The discussion of what good work exactly is and which factors affect the quality of work goes back to the 1970s [16]. The term good work is described and defined from different perspectives [17]. These perspectives cover specific organizational types. For instance, the type of work (e.g., production or knowledge work) should be considered. Also, the context factors, which influence good work relate to the professional area, e.g., software development [18].

In its 2020 annual report, the index of the German Trade Union Confederation (DGB) primarily refers to the common variants of mobile work or remote work. This focus is the result of the Covid-19 pandemic. In addition, possible context factors for good work are already mentioned in the report, some of which are adopted in this version for the survey or clustering. In summary, material security, development opportunities, recognition, sufficient work resources and low-stress activities make up good work [19]. The definition of the industrial union *IG Metall* overlaps in this respect, but supplements it with preventive and participation-oriented occupational health and safety as well as the sustainable handling of human performance [20]. This makes it clear that in addition to factors that can be shaped in the short term, sustainable aspects also play a role.

Finally, in connection with good work, the well-known methods of quality management are used. It is assumed here that quality development in an organization is successful if ethically based quality goals are set and the development of social skills is promoted.

In summary, it can be stated that good work can be understood as the framework conditions of a job that have a significant influence on the quality or quantity of the work results as well as on the quality of life of the individual.

### 2.2. *Gamification*

When playful elements with an aesthetic design are used in a non-game context, this is referred to as gamification [18]. Gamification typically contains a motivating mechanism that are used by users on a digital platform [21]. The trend towards gamification stems from the success of video games that have been reported in recent years. As a result of the digital transformation, gamification is increasingly used not only for private use, but also in commercial operations and other non-gaming contexts. In particular, gamification gained a lot of interest in the past years in the area of software engineering [22]. Gamification approaches are used to create incentives for employees to take certain actions. Entire workflows and processes can be adjusted accordingly. Gamification measures have a positive effect on motivation and strengthen the sense of togetherness within the team [18].

## 3. RESEARCH DESIGN

We performed a single case study at Otto with one embedded unit of analysis, which is the agile software development team under study. The study was designed and prepared based on the guidelines from Runeson and Hoest [24] and Yin [26]. We chose this approach because the topic of the study is timely and has not yet been sufficiently researched in this context. The selected research design allows to collect data based on flexible reactions of the agile software development team members during the research process [25].

We used three different research methods. In the first step, we performed a literature review to identify the context for good work. Based on our analysis, we created six cluster, which we used for the categorization of the identified factors from the literature. In the next step, we conducted a multipoint query with the agile software development team to identify the most relevant cluster of context factors. We used the results of the multipoint query to perform the third and final method. We performed an experiment with a gamification approach and collected the data in semi-structured interviews. We describe the specific research methods, the data collection and analysis in the following subsections.

### 3.1. *Clustering the context factors*

As described above, we performed a multipoint query using the tool Miro in order to get an understanding which of the clusters is of high importance from the agile software development teams' perspective. An overview of the cluster is available in Appendix A.1. The team voted for the health cluster. The health cluster consists of the following context factors: Mental stress, physical strain, leisure time and promotion of personal skills. As a result, we developed an experiment using a gamification approach for health-promoting measures in the context of remote work.

### 3.2. *Experiment*

In order to investigate the context factors for the health cluster, we decided to prepare and perform an experiment using a gamification approach with the agile software development team. To conduct the experiment remotely, we developed a gamification board with the virtual collaboration tool Miro.

At the end of an experiment week, the final weekly winner is determined and awarded a certificate in the daily meeting of the agile software development team. The activities are entered manually and independently by the agile software development team members, while the agile coach and lead of the team are responsible for the evaluation. To remind them to use the board and to exercise regularly, circular emails are sent to everyone involved twice a day with a short motivational text and a link to an approximately 5-minute exercise unit.

In addition, a mood barometer developed for the experiment is used daily in Miro in the team. The goal is to provide an opportunity for team reflection and to give the team leader an overview of potential dissatisfaction and stress in the team's working environment.

The experiment was conducted during two weeks in May 2021.

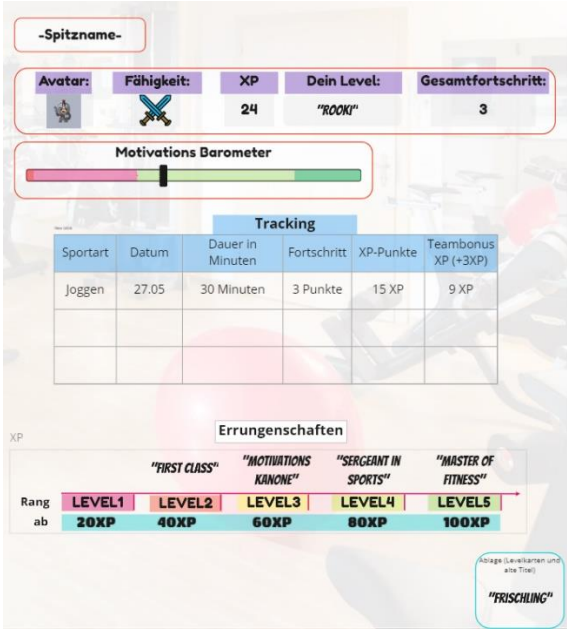

**Fig. 2**. Part of the gamification board

We selected this approach because no comparable experiments have been conducted in the context of our study. As shown in Figure 2, various game elements such as a level bar, ranks, skills and a motivation barometer are added to the tracking table. The boards of the individual subjects are in a common view, which means that the progress of the subjects is visible to everyone at all times. The leaderboard is updated on a daily basis to allow for constant comparison.

### 3.3. *Data Collection*

A semi-structured interview guideline with open questions was developed for the interviews, which enables a flexible reaction to the test persons' statements during the interviews. The interview guideline is available in Appendix A.2. To conduct the interviews, we performed video calls using Microsoft Teams. The interviews were performed in teams of two researchers. One researcher was responsible for conducting the interview and the other for documenting the answers. The interviewees are the members of the agile software development teams who voluntarily took part in the experiment. In total, we conducted eight interviews over a period of two weeks in May 2021.

### 3.4. *Data Analysis*

As a method for analyzing the interview data, qualitative content analysis according to Mayring [25] is used in this work. The summary content analysis is chosen inductively as the procedure since it is about exploring a new phenomenon and the categories are not explicitly defined in advance [27]. In the course of the summary, the individual interviews are summarized and the material is reduced to the research topic [25] with the aim of identifying commonalities in the interview statements and consequently of a high gain in knowledge for the derived recommendations for action achieve. The interview data were analyzed by the researcher team that did not conduct the interview itself in order to reduce the risk of bias.

## 4. RESULTS

In this Section, the results of the focus groups are presented and analyzed based on the six categories of motivation, social conversion, competitive spirit, implementation, self-organization, and well-being. The categories were identified in the course of structuring the interview statements and described in the following section. The descriptions of the dimensions are based on the statements of the subjects in the interviews conducted.

### 4.1. *Motivation*

During the remote work, the subjects did less sports overall than usual, although there was certainly the possibility to do so. This is due to the lack of motivation for physical activities that have arisen as a result of the pandemic. Despite existing health awareness, self-organized integration of exercise into everyday life is becoming increasingly difficult.

### 4.2. *Social conversion*

Social exchange suffers in a remote work setting. The willingness to communicate virtually is low because time is already spent at the work-station during work. Accordingly, face-to-face communication is lacking. However, there are also positive facets, such as the use of new communication and collaboration tools, which results in easier communication.

### 4.3. *Competitive spirit*

As presented in Section 2.2, the competition in an open gamification context represents a motivator, which may lead to an increased individual sports level. In particular, the successes of the other team members and the conscious examination of the topic increase motivation. On the other hand, a challenging approach is not a motivation for every employee. The sporting intentions and the comparison by the gamification board had a negative effect on individual team members, since the sport is perceived as an additional task.

### 4.4. *Implementation of a gamification approach*

The subjects rated the basic structure of the experiment as positive, intuitive and understandable. Only the distribution of the points and XP points is not comprehensible for all team members. In addition, the time factor is of importance, since the majority of the team members stated that they did not find any time for breaks and tracking to perform physical activities during their working time. The elements of gamification of the experiment are perceived as positive and the possibility of being able to climb levels increases motivation.

Some of the daily mails are perceived as spam by the team members, since the times of the reminders do not coincide with the break times due to the individual work structure. In addition, the e-mails are often deleted or automatically moved to folders in order to keep an overview of their tasks in the daily e-mails. However, the idea of remembering, in general, is rated as positive and gives an impetus to think more about physical activities, breaks and airing.

The mood barometer is predominantly perceived as positive. The evaluation showed that the team members were more motivated at the beginning of the week and in the morning than at the end of the week and in the afternoon. The team members consider it important to observe the mood in the team in order to be able to actively respond to the mood of their colleagues and to promote considerate cooperation. However, the lack of anonymity created a barrier to sharing negative moods.

### 4.5. *Individual self-organization*

It is not easy for team members with children to reconcile work and childcare, as there is a constant change of context between work and private life. In addition, work processes in remote work have to be reorganized. Although there is more time due to the elimination of commuting, it cannot be used effectively.

### 4.6. *Well-being*

Physical activities can help reduce stress and improve alertness. However, several team members point to the aspect that sport in front of colleagues causes discomfort. Accordingly, health promotion by the employer could also have negative effects.

## 5. PRACTICAL IMPLICATIONS

In the long term, the integration of the physical activity of the employees into everyday working life should become an integral part of the remote work culture. In order to establish this effectively, increased awareness of the importance of breaks and rest time slots is necessary. In addition, regular reminders to employees to keep working hours free are necessary. Similarly, adjusting the scheduling culture, in the form of fixed buffer times between appointments, can support to relaxing the workday.

To ensure a healthy level of physical activity for employees during working hours, we recommend establishing short breaks for physical activity as part of working hours. For this, employees should be given an appropriate time frame within their working hours as a tolerated break for physical activities, since the regular break times are too short, especially in the remote work context, to take adequate breaks and physical activities. It's possible that this could result in long-term benefits for businesses in the form of fewer cases of illness.

In order to promote social interaction among employees, it is recommended to arrange appointments within the team for which employees can voluntarily sign up on a list to participate in a joint virtual coffee break. These appointments can take place on a weekly or monthly basis or run parallel to iteration rhythms. Another option is a regular face-to-face meeting for team actions, such as joint walks.

To increase the added value of gamification, it is recommended to increase the level of automation to minimize the personal effort required to automatically track and update scores and tables. In this context, it is also recommended to use a different tool, as Miro is often seen as a work tool and cannot meet the requirements of employees due to its lack of anonymity and privacy. Particularly concerning mutual motivation in a team context, a solution should be chosen that makes it possible to motivate each other in real time without having to forgo anonymity.

Reminders for physical activity and airing should also be subscribable on a voluntary basis, with the possibility to choose the reminder times independently. In addition, the use of other communication channels than e-mail is recommended, as these may be overseen or ignored. One way to increase the exchange about physical and mental well-being in the team is to establish an anonymous mood barometer as part of agile practices such as retrospectives or daily stand-ups. It is also conceivable to develop new agile practices for this purpose that bring this facet to the fore.

## 6. THREATS TO VALIDITY

Although we designed our study based on the Runeson and Hoest guidelines, there are several limitations to consider.

*Construct validity:* A risk for construct validity is not including relevant and related literature when designing a case study. To counteract this risk, we searched various databases for relevant literature, deliberately including gray literature. This is a well-known approach for novel, hitherto little-researched topics [28]. We conducted all data (interviews and parts of the experiments) during the working hours of the team members in order to minimize potential risks of bias such as fatigue. Furthermore, we made sure that we only use tools for data collection (MS Teams and Miro) that were already known to the team members.

*Internal validity:* In order to avoid the risk of internal validity threats, we took several measures. First, we designed the interview guideline as semi-structured including non-leading questions. This approach allows as to go in-depth whenever an interviewee point to new or unexpected directions. The interviewees and the researchers did not know each other personally. However, we were not able to record and transcript the interview data due to data security issues. This limitation is covered by performing the interviews in pairs. We conducted all interviewers with at least two researchers. One performing the interview and the other observing and protocolling it.

*External validity:* The external validity could be optimized by adding more agile software development teams to our study. It is also worth to mention, that it would be interesting to add other cases from different business areas, companies or countries to this study. It could be interesting for a thorough analysis as the remote work setting affects many companies, teams and people around the globe.

## 7. CONCLUSION AND FUTURE WORK

The aim of our study was to examine which context factors are particularly relevant for good work at Otto in an agile software development team in current remote work setting and how these can be designed accordingly. The results of our study show that team members attach the greatest importance to the area of health. Our results show that external incentives such as the use of gamification elements motivate the majority of the team members to improve their physical activities and awareness of health well-being. After conducting the interviews, the research question can be answered as follows: For the employees of a software development team at Otto, the area of health is of the greatest importance. This area can be designed accordingly through gamification and leads to at least short-term positive effects. It should be mentioned here that gamification has not led to an increase in movement for some employees. Our work serves as a basis for further research. It would be interesting how agile software development teams from other contexts (e.g. industries, companies and countries) react to our gamification experiment and which context factors for good work they focus on. Since the comprehensive introduction of remote work also affects other business areas, we also recommend considering other (software) process models or roles in future research in this area.

## APPENDIX

A.1: The overview of the cluster is available at the following link:

https://sync.academiccloud.de/index.php/s/XJ5hjRK1Gfjf2LG

A.2: The interview guideline is available at the the following link:

https://sync.academiccloud.de/index.php/s/G7AFDIjf3QGf3JF